\documentclass[12pt]{article}

\usepackage{amsfonts,latexsym,eucal,color,graphicx,epsfig,amssymb,amsmath}

\newcommand{\be}{\begin{eqnarray}}
\newcommand{\ee}{\end{eqnarray}}
\newcommand{\beq}{\begin{eqnarray}}
\newcommand{\eeq}{\end{eqnarray}}
\newcommand{\pd}{\partial}

\newcommand{\dalm}{\kern1pt\vbox{\hrule height 0.9pt\hbox{\vrule width 0.9pt\hskip 2.5pt\vbox{\vskip 5.5pt}\hskip 3pt\vrule width 0.3pt}\hrule height 0.3pt}\kern1pt}
\newcommand{\ie}{{\it i.e.}}
\newcommand{\eg}{{\it e.g.}}

\makeatletter

\@addtoreset{equation}{section}
\makeatletter

\begin{document}

\thispagestyle{empty}
\begin{titlepage}

\begin{flushright}
April 2, 2012
\end{flushright}
\vspace{3cm}
\begin{center}
{\Large {\bf Information metric from a linear sigma model}}\\
\vspace{1.5cm}
{\bf Umpei Miyamoto \quad\quad Shigeaki Yahikozawa}\\
\vspace{.5cm} 
{\it Department of Physics, Rikkyo University, Tokyo 171-8501, Japan}\\
\vspace{.5cm}
{\tt umpei@rikkyo.ac.jp \quad\quad yahiko@rikkyo.ac.jp}\\
\vspace{1cm}
\end{center}

\begin{abstract}
The idea that a spacetime metric emerges as a Fisher-Rao `information metric' of instanton moduli space has been examined in several field theories such as the Yang-Mills theories and nonlinear sigma models. In this brief paper, we report that the flat Euclidean or Minkowskian metric, rather than an anti-de Sitter metric that generically emerges from instanton moduli spaces, can be obtained as the Fisher-Rao metric from a non-trivial solution of the massive Klein-Gordon field (a linear sigma model). This realization of the flat space from the simple field theory would be useful to investigate the ideas that relate the spacetime geometry with the information geometry. 
\end{abstract}

\end{titlepage}

\section{Introduction}

In the information geometry~\cite{AmariNagaoka,MurrayRice}, a Fisher (or Fisher-Rao) metric is defined on a statistical manifold. It may be written as
\be
	g_{ab}(\theta)
	=
	\int d^Dx P(x;\theta) \pd_{a} \ln P(x;\theta) \pd_{b} \ln P(x;\theta),
\label{Fisher}
\ee
where $x^\mu = (x^0, x^1, \ldots, x^{D-1})$ is a set of random variables and $\theta^a = (\theta^0, \theta^1, \ldots, \theta^{N-1})$ is a set of statistical parameters of a probability distribution $P(x;\theta)$. The distance given by $ |g_{ab}(\theta) d\theta^a d\theta^b |^{1/2}$ is called the Rao distance and used as a measure of difference between two probability distributions. As usual, the probability density function $P(x;\theta)$ should be normalized
\be
	\int d^D x P(x;\theta) = 1,
\label{norm}
\ee
while for an arbitrary function of random variables $f(x)$, its expectation value $E[f]$ and variance $V[f]$, being functions of parameters $\theta$, are given by
\begin{align}
\begin{split}
	E[f](\theta)
	&:=
	\int d^D x f(x) P(x;\theta),
\\
	V[f](\theta)
	&:=
	E[ ( f - E[f] )^2 ]
	=
	E[ f^2 ] - E[f]^2.
\end{split}
\end{align}

While geometrical approaches have many advantages in statistical physics (see, \eg, \cite{Ruppeiner:1995zz}), the above information metric seems just a mathematical notion/tool and far from the measure of distance in physical space. 
However, the notions in the information geometry are attractive for physicists because at the Planckian regime ($ \lesssim 10^{-33}\;{\rm cm}$), where quantum gravitational effects dominate, the notion of spacetime points and distance among spacetime points could be calculable statistically rather than deterministically. Thus, the notions and techniques in the information geometry could be useful to describe spacetimes at the Planckian regime.\footnote{Although we have in mind this expectation, we do not insist necessarily that the physics in this paper involve a quantum gravitational effect. In fact, there appears no specific scale corresponding to the Planck length in this paper.} This expectation is somewhat similar to that many physicists expect the non-commutative geometry~\cite{Connes:1994yd} to be useful in describing quantum geometry.

One might say that the above expectation is partially realized in the AdS/CFT (anti-de Sitter/conformal field theory) correspondence~\cite{Maldacena:1997re} (see, \eg, \cite{Aharony:1999ti} for a review), which asserts the equivalence of a gravitational theory (\ie, the geometry of spacetime~\cite{Hawking:1973uf}) and a quantum field theory~\cite{Peskin:1995ev} at the boundary of spacetime, being lower dimensional. Namely, it was shown in reference~\cite{Blau:2001gj} that the bulk AdS geometry, where the gravitational theory is defined, can emerge as the information metric of the instanton moduli space of the boundary Yang-Mills theory. In such applications of information geometry to field theories, spacetime coordinates and parameters in a solution of field theory (\eg, the moduli of instanton solution for example) are identified with the random variables $x^\mu$ and statistical parameters $\theta^a$, respectively, while a Lagrangian density plays the role of probability distribution, which was first suggested by Hitchin~\cite{Hitchin}. Surprisingly, a black-hole geometry~\cite{Hawking:1973uf} being asymptotically AdS can also be obtained as the Fisher metric from the Yang-Mills instantons~\cite{Rey:2005cn,Rey:2006bz}.\footnote{Their results are not only interesting in that a highly non-trivial geometrical object, the black hole, emerges but also it happens beyond the strong-coupling regime where the original AdS/CFT correspondence~\cite{Maldacena:1997re} was proposed.}

While the Fisher metric of the instanton moduli spaces has been examined in the AdS/CFT context, it would be of great interests to understand from more general point of view what kind of field theories (and their solutions) can produce fundamental geometries like a flat space, an AdS space, and a black hole as the information geometry. Regarding this point, one of the present authors (S.Y.) illustrated that the AdS space can be obtained from the instanton solutions of {\it nonlinear sigma models}~\cite{Yahikozawa:2003ij}, which was an example where the non-trivial Fisher metric is obtained from a field theory not being Abelian nor non-Abelian Yang-Mills theories. In this paper we would like to expand this development to a much simpler but non-trivial example, where an Euclidean or pseudo-Euclidean flat space emerges from a {\it linear sigma model} (precisely speaking, a massive Klein-Gordon scalar field), being one of simplest relativistic field theories.

To find what kind of probability distribution produces the flat space, it seems essential to notice that one can always find Cartesian coordinates for a flat space, in which the metric components are constant $g_{ab}(\theta) = g_{ab}(0)$. For simplicity let us consider a class of systems in which the numbers of spacetime dimensions (as the random variables) and the parameters coincide $D=N$. Moreover, let us suppose that the distribution $P(x;\theta)$ takes the form of $P(x;\theta) = P(x-\theta)$.\footnote{Then, both the Greek ($\mu,\nu,\ldots$) and Latin ($a,b,\ldots$) indices can be used to specify the components of random variables $x$ and parameters $\theta$.} With these assumptions, one can show that the information metric $g_{ab}(\theta)$ has no dependence on $\theta$:
\begin{align}
\begin{split}
	g_{ab} ( \theta)
	&=
	\int d^D x \frac{ \pd_{\theta^a} P(x-\theta) \pd_{\theta^b} P(x-\theta) }{ P(x-\theta) }
\\
	&=
	\int d^D x \frac{ \pd_{x^a} P(x) \pd_{x^b} P(x) }{ P(x) }
	= g_{ab}(0).
\label{flatness}
\end{split}
\end{align}
Here, the integration range is assumed to be whole of $ {\mathbb R}^D = (-\infty,\infty) \times \cdots \times (-\infty,\infty) $, and this infiniteness is necessary for the second equality in (\ref{flatness}) to hold. Given this observation, we will see in the next section that a solution in the linear sigma model indeed presents an example of flat geometry whose metric is constant, definite, and non-degenerate.

\section{Emergent flat space from the Klein-Gordon field }

\subsection{Massive scalar field}

The Lagrangian density and action integral of the massive Klein-Gordon scalar field $\phi$, one of the linear sigma models, are given by
\be
	\mathcal{L}(x)
	=
	\frac12 \left( \eta^{\mu\nu} \pd_\mu \phi \pd_\nu \phi -  m^2 \phi^2 \right),
\;\;\;
	S[\phi] 
	=
	\int d^D x \mathcal{L}(x),
\label{lagrangean}
\ee
where $\eta^{\mu\nu} = \eta_{\mu\nu} = {\rm diag}(1,-1,\ldots,-1)$ is a $D$-dimensional Minkowski (pseudo-Euclidean) metric and $m$ is a constant corresponding to the mass of scalar field. The equation of motion, obtained by setting zero the variation of action $\delta S/\delta \phi = 0$, is
\be
	( \Box + m^2 ) \phi = 0,
\;\;\;
	\Box := \eta^{\mu\nu} \pd_\mu \pd_\nu.
\label{eom}
\ee

One can easily show that the following function solves the equation of motion (\ref{eom}),
\be
	\phi = A \exp\left( - \frac{ m }{\sqrt{D-2}} |x|\right) ,
\;\;\;
	|x| := \sum_{\mu=0}^{D-1} |x^\mu|,
\;\;\;
	D \geq 3.
\label{sol}
\ee
When $D=2$, this type of solution does not exist for the massive case ($m > 0$). At this point, $A$ is an arbitrary constant since the equation of motion (\ref{eom}) is linear. However, we will fix this constant later to make the model allow appropriate statistical interpretations.

\subsection{Information metric}

Hitchin~\cite{Hitchin} proposed to adopt the square of field strength $F^2$ of Yang-Mills theory as a probability distribution. A natural generalization of Hitchin's proposal is to regard the minus on-shell Lagrangian density (\ie, the Lagrangian density evaluated on a solution) as the probability distribution. Thus, we take
\begin{align}
\begin{split}
	P(x-\theta) 
	&:= - \mathcal{L}(x-\theta)
\\
	&= A^2 m^2 \exp\left( - \frac{ 2 m }{\sqrt{D-2} } |x-\theta|  \right),
\end{split}
\label{onshell}
\end{align}
where we have evaluated the Lagrangian (\ref{lagrangean}) on solution (\ref{sol}).\footnote{The solution~(\ref{sol}) is $C^0$ (\ie, not differentiable) at $x^\mu = \theta^\mu $ but the Lagrangian and its integration over the whole spacetime can be evaluated appropriately nonetheless.} In order to be compatible with this identification (\ref{onshell}), the constant $A$ should be fixed by the normalization condition (\ref{norm}),
\be
	A = \frac{m^{(D-2)/2}}{ (D-2)^{D/4} }.
\label{A}
\ee

\begin{figure}[t!]
	\begin{center}
			\includegraphics[width=8cm]{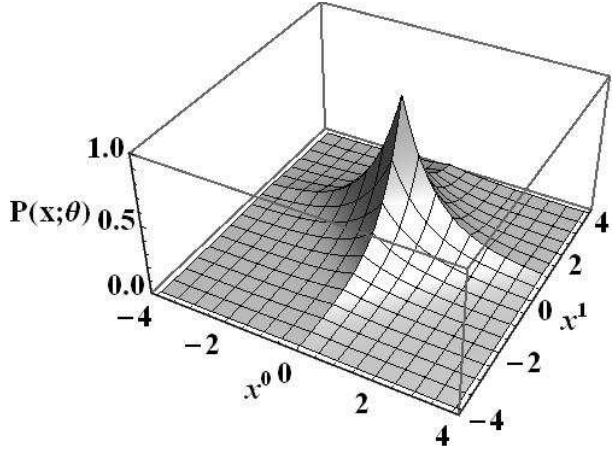}
	\caption{{\it Probability distribution given by equations (\ref{onshell}) and (\ref{A}) for the $D=3$ case, where $x^2$-coordinate is suppressed. The statistical parameters and mass set as $\theta^0=0$, $\theta^1 = 1$, and $m=1$. One can see that the distribution is localized with the width $ \sim 1/m = 1$ around $(x^0,x^1)=(\theta^0,\theta^1)$ .}}
	\label{fg:prob}
	\end{center}
\end{figure}

Although we have introduced the statistical parameter $\theta^a$ just as the `shift' of coordinates at the end of Introduction, one can associate physical/geometrical meaning to it and the mass parameter $m$ by evaluating the expectation value and variance of variable $x^\mu$ as follow,
\begin{align}
\begin{split}
	E[x^\mu](\theta)
	&=
	- \int d^D x x^\mu \mathcal{L}(x-\theta)
\\
	&=
	E[x^\mu] (0) - \theta^\mu \int d^D x \; \mathcal{L}(x) 
	= \theta^\mu,
\end{split}
\end{align}
where we have used $ E[x^\mu] (0) = 0 $ and normalization condition (\ref{norm}). That is, $\theta^\mu$ is the expectation value of $x^\mu$ itself, which just reflects that the distribution is centered at $x^\mu = \theta^\mu$. On the other hand, one can see that the inverse of mass $m^{-1}$ corresponds to the standard deviation (\ie, the square root of variance) of $x^\mu$,
\begin{align}
\begin{split}
	V[x^\mu](\theta)
	&=
	-\int d^D x (x^\mu)^2 \mathcal{L}(x-\theta) - (\theta^\mu)^2
\\
	&=
	E[(x^\mu)^2](0)
	=
	\frac{1}{2 (D-2)^{(D-6)/2} m^2},
\end{split}
\end{align}
which just represents that the distribution is localized with the scale $m^{-1}$.\footnote{In terms of quantum mechanics, this length scale is nothing but the Compton wave length $\hbar/mc$.} See figure \ref{fg:prob}.

Now, we are ready to calculate the Fisher metric (\ref{Fisher}) using the probability density~(\ref{onshell}). This probability distribution is in the cases where the relation (\ref{flatness}) holds. Thus, the information metric can be evaluated after setting $\theta = 0$:
\be
	g_{ab}
	=
	\frac{ 4m^{D+2} }{ (D-2)^{ (D+2)/2 } }
	\int d^D x \; {\rm sgn}(x^a) {\rm sgn}(x^b)
	\exp\left( - \frac{ 2 m }{\sqrt{D-2} } |x|  \right),
\ee
where ${\rm sgn}(\xi) := \xi/|\xi|$ is the sign function.  
This integration can be evaluated to yield
\be
	g_{ab}
	=
	\frac{ 4 m^2 }{ D-2 } \delta_{ab}.
\label{flatMetric}
\ee
Thus, we have obtained the flat Euclidean metric as the Fisher metric from the linear sigma model. By scaling the coordinates as $ \theta^a \to \bar{\theta}^a = ( 2m/\sqrt{D-2} ) \theta^a $, the metric can be normalized $g_{ab} \to \bar{g}_{ab} = \delta_{ab}$. Furthermore, carrying out the Wick rotation $ \theta^a \to {\rm i} \theta^a $ (${\rm i}^2 = -1$) in necessary components one can even get a metric of pseudo-Euclidean space, \eg, $g_{ab} \to \eta_{ab}$.

\section{Concluding remarks}

We have obtained the flat metric (\ref{flatMetric}) as the Fisher-Rao information metric~(\ref{Fisher}) from the linear sigma model (or massive Klein-Gordon scalar model), which is defined by the Lagrangian density (\ref{lagrangean}). While it is well known that the (Euclidean) AdS geometry often emerges in the instanton moduli spaces~\cite{Blau:2001gj,Rey:2005cn,Rey:2006bz,Yahikozawa:2003ij}, we have found that the flat metric emerges from the {\it non-trivial} solution (\ref{sol}) in the {\it simple} field theory (\ref{lagrangean}).

We would like to stress from several points of view the non-triviality of the emergence of flat metric from a field theory. One point is that the general argument presented at the end of Introduction does not necessary imply the existence of a flat metric, that must be {\it non-degenerate} and {\it definite} by the definition of a metric. Another point is that we have obtained the flat metric from {\it a field theory}. Namely, for example, it is easy to check that a Gaussian distribution $ P(x-\theta) = \pi^{-D/2} \exp [ -(x-\theta)^2 ] $, where $ x^2 := \sum_{\mu=0}^{D-1} (x^\mu)^2 $, yields a flat metric $ g_{ab} = 2\delta_{ab} $. It is highly non-trivial, however, to find a physically reasonable relativistic field theory and its solution whose on-shell Lagrangian yields such a Gaussian distribution. In fact, the field theory~(\ref{lagrangean}) and its solution discussed in this paper~(\ref{sol}) were obtained after much trial and error.
Finally, it is added that the emergence is non-trivial from the viewpoint that the solution (\ref{sol}) is {\it anisotropic}. Namely, one may be able to argue (or possibly can prove) that the information metric is proportional to the Kronecker delta, which is the simplest isotropic rank-2 tensor, once the probability distribution is assumed to have rotational symmetry around a point (\eg, $x^\mu = \theta^\mu$ in the present case). However, the solution (\ref{sol}) has no rotational symmetry but has only discrete reflection symmetries, and therefore the probability distribution is anisotropic.

Finally, let us mention that there have been attempts to derive an equation of motion that the Fisher information metric obeys in the literature (see, \eg, \cite{Frieden,CalmetCalmetPRE2005}).\footnote{Precisely speaking, these papers derive the equation of motion for a wave function whose counterpart in the present paper is $\phi(x)$~\cite{Frieden}, and the equation of motion for the probability distribution $P(x;\theta)$~\cite{CalmetCalmetPRE2005}, rather than that for the information metric itself. The present authors thank an anonymous referee for pointing out this point.} While our approach is different from those in the literature, it is nice to see that the flat metric obtained in this paper is the most fundamental vacuum solution to the Einstein field equation~\cite{Hawking:1973uf}, just like that the AdS metric is a fundamental solution to the Einstein equation with a negative cosmological constant. As the Einstein equation is one of the simplest gravitational equations satisfying the fundamental physical requirements such as the equivalence principle and general covariance, one could expect the Fisher metric to obey an equation of motion similar to the Einstein equation in an appropriate limit. The flat metric obtained in this paper from the simple field theory would be a footstep to investigate such a possibility.

\subsection*{Acknowledgments}

We would like to thank T.~Kuroki and anonymous referees for useful discussions and comments.
This work is supported by Research Center for Measurement in Advanced Science in Rikkyo University, by Rikkyo University SFR (Special Fund for Research), and by the Grant-in-Aid for Scientific Research Fund of the Ministry of Education, Culture, Sports, Science and Technology, Japan (Young Scientists (B) 22740176).



\end{document}